\documentclass[showpacs,amssymb,preprint,preprintnumbers,nofootinbib,superscriptaddress]{revtex4}
\usepackage{epsf,epsfig,graphics,graphicx}
\usepackage{verbatim,color,ulem}
\usepackage{xcolor}
\usepackage{subcaption}
\begin{document}

\title{Open quantum system approach to the Unruh-DeWitt detector in impulsive plane wave spacetimes}
\author{Hing-Tong Cho}
\email[Email: ]{htcho@mail.tku.edu.tw}
\affiliation{Department of Physics, Tamkang University, Tamsui, New Taipei City 25137, Taiwan}

\date{\today}

\begin{abstract}
In this paper we employ the open quantum system framework with the influence functional formalism to non-perturbatively analyze the response of an Unruh-DeWitt detector modeled as a harmonic oscillator which interacts with a massless scalar field in impulsive plane wave spacetimes. Subtracting the Minkowski results, we obtain the expectation values for $\langle Q^2 \rangle$, $\langle P^2 \rangle$, and $\langle \{Q,P\} \rangle$, along with transition probabilities $P_{0\rightarrow 1}$ from ground state to the first excited state of the detector due to the influence of the wave. Explicit calculations are performed for both a pure gravitational wave (vanishing Ricci tensor) and a null electromagnetic wave (vanishing Weyl tensor). In both scenarios, the wave suppresses excitation transitions. Since our approach is non-perturbative, we are able to consider cases with both weak and strong coupling constants.

\end{abstract}

\pacs{
}
\maketitle

\section{Introduction}








Through the Penrose limiting procedure \cite{Blau11}, one can zoom infinitely close to a null geodesic by applying appropriate boosts and scalings, revealing that the local geometry simplifies to a plane wave spacetime. As Penrose \cite{Penrose76} famously observed, "Any spacetime has a plane wave as a limit." This geometric framework offers a powerful non-perturbative avenue for investigating quantum field effects within general curved backgrounds. For example, Shore and collaborators \cite{HS08,HSS09} utilized this limiting technique to calculate the one-fermion-loop correction to the photonic refractive index and dispersion relation in quantum electrodynamics. Their analysis successfully proved that photon propagation in QED on curved backgrounds remains strictly causal and dispersive across the entire frequency spectrum.

In this work, we focus specifically on impulsive plane waves whose profile is modeled by a Dirac $\delta$-function. The spacetime geometry is remarkably clean: the region strictly before the arrival of the wave and the region following its departure are simply flat Minkowski spacetime. For this reason exact mode functions and their corresponding two-point correlation functions can be found. In our earlier work \cite{Cho23}, we explored several classical and quantum features of this background, paying particular attention to the evaluation of the Wightman function, which plays a central role in our subsequent derivations in this paper. Here, to deepen our understanding of quantum dynamics in this background, we examine the response of an Unruh-DeWitt (UD) detector \cite{Unruh76,DeWitt79,BD82} subject to the sudden passage of an impulsive wave, extending key insights developed in recent studies by Gray et al. \cite{GKMTT21} as well as Pitelli and Mosna \cite{PM24}.

UD detectors serve as idealized quantum mechanical systems, such as harmonic oscillators or localized atoms with discrete internal energy states, designed to perform local measurements through direct coupling to quantum fields in general curved spacetimes. This conceptual detector model was originally introduced to bypass the fundamental ambiguity of defining a global ``particle" state in general relativity, where spacetimes frequently lack asymptotically flat regions or globally timelike Killing vector fields. Because standard particle definitions depend heavily on global spacetime symmetries that are rarely available a priori \cite{BD82}, quantum transitions between internal detector states offer a physically meaningful, local measurement which signifies the presence of particles. Consequently, UD detectors have become standard diagnostic tools for probing phenomena such as the Unruh effect for accelerating observers \cite{BD82,CHM08} and Hawking radiation in black hole backgrounds \cite{Hawking75,Wald01}.

Although detector excitation rates are traditionally calculated using lowest-order time-dependent perturbation theory \cite{Unruh76,DeWitt79}, the underlying interaction between the detector and the surrounding quantum field can be treated exactly within an open quantum system paradigm. Exemplified by the quantum Brownian motion framework \cite{Schwinger61,Keldysh64}, this formulation uses reduced density matrices and Feynman-Vernon influence functionals \cite{FV63} to incorporate field-induced backreaction effects self-consistently. To describe open quantum dynamics, one employs the in-in or Schwinger-Keldysh formalism, where path-integral trajectories progress forward and backward along a closed time path \cite{ZSHY85,HPZ92,HPZ93}. Integrating out the environmental field degrees of freedom generates effective dissipation and noise terms, where quantum field fluctuations manifest as a stochastic force driving a generalized Langevin equation of motion \cite{CKW04}. Developed comprehensively by Hu and collaborators \cite{JH00,JH02,HH19,GH05,GHY06,LH07,HLL12,HH19a}, this non-perturbative framework allows us to evaluate the complete response of a UD detector to an impulsive plane wave beyond the perturbative weak-coupling limit. In particular, we shall follow closely the method put forth in \cite{HH19a}.

This paper is organized as follows. Section II presents the open quantum system framework used to evaluate the response of a UD detector due to its interaction with the quantum field in a general curved spacetime in a non-perturbative manner. We derive the reduced density matrix of the detector by integrating out the environmental quantum field degrees of freedom, subsequently using this density matrix to compute key operator expectation values and the transition probabilities of the detector from the ground state to the first excited state. In Section III, we apply this formalism to the background spacetime of an impulsive plane wave. We place particular emphasis on the analysis of the homogeneous equations with dissipation, whose solutions are crucial in our later derivations. By subtracting the Minkowski contribution, we are able to isolate the effect of the impulsive wave on the transition probabilities of the detector. In Section IV, two explicit examples, one for a gravitational wave and the other for an electromagnetic one, are considered, examining cases with different coupling constants and strengths of the wave. Finally, Section V provides concluding remarks and discusses avenues for future research.

\section{Unruh-DeWitt detector: Open quantum system approach}

Consider the UD detector with a prescribed trajectory $z^{\mu}(\tau)$ with $\tau$ being its proper time. The internal degree of freedom $Q$ of the detector is assumed to be described by a harmonic oscillator with the action
\begin{eqnarray}
    S_{D}[Q]=\int\,d\tau\,\left(\frac{m}{2}\right)\left[(\partial_{\tau}Q)^{2}-\omega_{0}^{2}\,Q^{2}\right]
\end{eqnarray}
with mass $m$ and bare frequency $\omega_{0}$. $Q$ couples to a massless scalar field $\Phi$ in the background spacetime spacetime with metric $g_{\mu\nu}$. The free action of $\Phi$ is given by
\begin{eqnarray}
    S_{F}[\Phi]=\int\,d^{4}x\sqrt{-g}\left(\frac{1}{2}\,\partial_{\mu}\Phi\,\partial^{\mu}\Phi\right)
\end{eqnarray}
The interaction action between $Q(\tau)$ and $\Phi(x)$ is taken to be bilinear,
\begin{eqnarray}
    S_{I}[Q,\Phi]=\lambda\int d\tau\int d^{4}x\,Q(\tau)\,\Phi(x)\,\delta^{4}(x^{\mu}-z^{\mu}(\tau))
\end{eqnarray}
The total action therefore takes the form
\begin{eqnarray}
    S[Q,\Phi]=S_{D}[Q]+S_{F}[\Phi]+S_{I}[Q,\Phi]
\end{eqnarray}

\subsection{Density matrix and evolution operator}
In the UD detector theory one is interested in obtaining the transition probabilities between various energy levels of $Q$ under the influence of the quantum field $\Phi$ in the corresponding background spacetime. To achieve that we study the reduced density matrix $\rho_{R}$ which describes the quantum state of $Q$ after integrating out the influence of the quantum field. The time evolution of $\rho_{R}$ is specified by
\begin{eqnarray}\label{rhoR}
    \rho_{R}(Q_{+f},Q_{-f},\tau_{f})=\int_{-\infty}^{\infty}dQ_{+i}\int_{-\infty}^{\infty}dQ_{-i}\ {\cal J}_{R}(Q_{+f},Q_{-f},\tau_f|Q_{+ i},Q_{-i},\tau_{i})\ \rho_{R}(Q_{+i},Q_{-i},\tau_i)
\end{eqnarray}
where ${\cal J}_{R}$ is the evolution operator of the reduced density matrix. In the open quantum system approach, in particular via the path-integral formalism, this evolution operator can be expressed as \cite{HPZ92,HM94}
\begin{eqnarray}\label{ctp}
    {\cal J}_{R}(Q_{+f},Q_{-f},\tau_{f}|Q_{+ i},Q_{-i},\tau_{i})=\int_{Q_{\pm
    i}}^{Q_{\pm f}}DQ_{\pm}\int_{CTP}D\Phi_{\pm}\,e^{iS[Q_{+},\Phi_{+}]-iS[Q_{-},\Phi_{-}]}
\end{eqnarray}
where $CTP$ represents the so-called ``closed-time-path" boundary condition for the path integrals over $\Phi_{\pm}$. Note that we have assumed that initially the detector and the field are uncorrelated to arrive at the expression in Eq.~(\ref{ctp}). Since the field $\Phi$ is at most quadratic in the action $S[Q,\Phi]$, the CTP path integrals can be evaluated exactly. The result is known as the influence action $S_{IF}$,
\begin{eqnarray}\label{SIF}
    e^{iS_{IF}[\Delta,\Sigma]}&=&\int_{CTP}D\Phi_{\pm}\,e^{i(S_{F}[Q_{+}]+S_{I}[Q_{+},\Phi_{+}])-i(S_{F}[Q_{-}]+S_{I}[Q_{+},\Phi_{+}])}\nonumber\\
    &=&e^{i\lambda^2\int_{\tau_i}^{\tau_f} ds\int_{\tau_i}^{\tau_f} ds'\,\Delta(s)D(s,s')\Sigma(s')-\frac{1}{2}\lambda^2\int_{\tau_i}^{\tau_f} ds\int_{\tau_i}^{\tau_f} ds'\,\Delta(s)N(s,s')\Delta(s')}
\end{eqnarray}
where $\Delta=Q_{+}-Q_{-}$ and $\Sigma=(Q_{+}+Q_{-})/2$. Here $D(s,s')$ is called the dissipation kernel which is given by the retarded Green function of the field $\Phi$ in the corresponding spacetime,
\begin{eqnarray}\label{disker}
    D(s,s')=G_{ret}(z(s),z(s'))=i\,\theta(z^{0}(s)-z^{0}(s'))\left\langle[\Phi(z(s)),\Phi(z(s'))]\right\rangle
\end{eqnarray}
and $N(s,s')$ is the noise kernel given by the Hadamard function of the field,
\begin{eqnarray}\label{noiker}
    N(s,s')=G^{(1)}(z(s),z(s'))=\frac{1}{2}\left\langle\{\Phi(z(s)),\Phi(z(s'))\}\right\rangle
\end{eqnarray}
Recall that $z(s)$ is the prescribed trajectory of the detector.

With the result of the influence action, the evolution operator can be written as
\begin{eqnarray}\label{intSD}
    {\cal J}_{R}(Q_{+f},Q_{-f},\tau_{f}|Q_{+ i},Q_{-i},\tau_{i})&=&\int_{Q_{\pm
    i}}^{Q_{\pm f}}DQ_{\pm}\,e^{i(S_{D}[Q_{+}]-S_{D}[Q_{-}]+S_{IF}[Q_{+},Q_{-}])}\nonumber\\
    &=&\int_{\Sigma_{i}}^{\Sigma_{f}}D\Sigma\int_{\Delta_{i}}^{\Delta_{f}}D\Delta\ e^{iS_{eff}[\Sigma,\Delta]}
\end{eqnarray}
where $S_{eff}$ is the effective action 
\begin{eqnarray}
S_{eff}[\Sigma,\Delta]=\int_{\tau_{i}}^{\tau_f}d\tau\,m\,[\partial_{\tau}\Sigma(\tau)\partial_{\tau}\Delta(\tau)-\omega_{0}^{2}\,\Sigma(\tau)\Delta(\tau)]+S_{IF}[\Delta,\Sigma]
\end{eqnarray}
To evaluate the path integrals over $\Delta$ and $\Sigma$, we deal with their corresponding boundary conditions by splitting the functions into \cite{HM94}
\begin{eqnarray}
    \Sigma(s)=\Sigma_{cl}(s)+\sigma(s)\hskip 20pt;\hskip 20pt \Delta(s)=\Delta_{cl}(s)+\delta(s)
\end{eqnarray}
where $\Sigma_{cl}$ and $\Delta_{cl}$ are solutions to the classical equations of motion,
\begin{eqnarray}
    &&\partial_{s}^{2}\Sigma_{cl}(s)+\omega_{0}^{2}\,\Sigma_{cl}(s)-\left(\frac{\lambda^{2}}{m}\right)\int_{\tau_{i}}^{\tau_f}ds'\,D(z(s),z(s'))\,\Sigma_{cl}(s')=0\label{cleq1}\\
    &&\partial_{s}^{2}\Delta_{cl}(s)+\omega_{0}^{2}\,\Delta_{cl}(s)-\left(\frac{\lambda^{2}}{m}\right)\int_{\tau_{i}}^{\tau_f}ds'\,\Delta_{cl}(s')\,D(z(s),z(s'))=0\label{cleq2}   
\end{eqnarray}
with the boundary conditions
\begin{eqnarray}
    \Sigma_{cl}(\tau_i)=\Sigma_{i}\hskip 20pt &&;\hskip 20pt \Delta_{cl}(\tau_{i})=\Delta_{i}\nonumber\\
    \Sigma_{cl}(\tau_f)=\Sigma_{f}\hskip 20pt &&;\hskip 20pt \Delta_{cl}(\tau_{f})=\Delta_{f}
\end{eqnarray}
On the other hand, $\sigma(s)$ and $\delta(s)$ are arbitrary functions with boundary conditions 
\begin{eqnarray}
    \sigma(\tau_{i})=\delta(\tau_{i})=0\hskip 20pt;\hskip 20pt \sigma(\tau_f)=\delta(\tau_f)=0
\end{eqnarray}
With this splitting, the path integrals of the evolution operator in Eq.~(\ref{intSD}) can be further separated as
\begin{eqnarray}\label{Jpre}
    &&{\cal J}_{R}(Q_{+f},Q_{-f},\tau_f|Q_{+ i},Q_{-i},\tau_{i})\nonumber\\
    &=&e^{iS_{eff}[\Sigma_{cl},\Delta_{cl}]}\int\,D\sigma\int\,D\delta\ e^{iS_{eff}[\sigma,\delta]-\lambda^2\int_{\tau_{i}}^{\tau_f}ds\int_{\tau_{i}}^{\tau_f}ds'\,\Delta_{cl}(s)N(s,s')\,\delta(s')}
\end{eqnarray}
where we have made use of the equations of motion in Eqs.~(\ref{cleq1}) and (\ref{cleq2}).

First, we consider the prefactor involving $S_{eff}[\Sigma_{cl},\Delta_{cl}]$.
Suppose that $u_{1}$ and $u_{2}$ are solutions to Eq.~(\ref{cleq1}) with boundary conditions
\begin{eqnarray}\label{u12bc}
    u_{1}(\tau_{i})=1=u_{2}(\tau_f)\hskip 20pt;\hskip 20pt u_{1}(\tau_f)=0=u_{2}(\tau_{i}),
\end{eqnarray}
while $v_{1}$ and $v_{2}$ are solutions to Eq.~(\ref{cleq2}) with boundary conditions
\begin{eqnarray}\label{v12bc}
    v_{1}(\tau_{i})=1=v_{2}(\tau_f)\hskip 20pt;\hskip 20pt v_{1}(\tau_f)=0=v_{2}(\tau_{i})
\end{eqnarray}
Using these two sets of fundamental solutions, we can write
\begin{eqnarray}
    \Sigma_{cl}(s)=\Sigma_{i}\,u_{1}(s)+\Sigma_{f}\,u_{2}(s)\\
    \Delta_{cl}(s)=\Delta_{i}v_{1}(s)+\Delta_{f}\,v_{2}(s)
\end{eqnarray}
Putting these into $S_{eff}[\Sigma_{cl},\Delta_{cl}]$, the prefactor in Eq.~(\ref{Jpre}) is simplified to
\begin{eqnarray}
    e^{iS_{eff}[\Sigma_{cl},\Delta_{cl}]}=e^{ib_{1}\Sigma_f\Delta_f -ib_{2}\Sigma_f\Delta_{i}+ib_{3}\Sigma_{i}\Delta_f-ib_{4}\Sigma_{i}\Delta_{i}}\,e^{-a_{11}\Delta_{i}^2-a_{12}\Delta_{i}\Delta_f-a_{22}\Delta_f^{2}}
\end{eqnarray}
where
\begin{eqnarray}
    &&b_{1}=m\left[\partial_{s}u_{2}\right]_{s=\tau_f}\hskip 20pt;\hskip 20pt b_{2}=m\left[\partial_{s}u_{2}\right]_{s=\tau_{i}}\nonumber\\
    &&b_{3}=m\left[\partial_{s}u_{1}\right]_{s=\tau_f}\hskip 20pt;\hskip 20pt b_{4}=m\left[\partial_{s}u_{1}\right]_{s=\tau_{i}}\label{bcoeff1}\\
    &&a_{ij}=\frac{\lambda^{2}}{1+\delta_{ij}}\int_{\tau_{i}}^{\tau_f}ds\int_{\tau_{i}}^{\tau_f}ds'\,v_{i}(s)N(s,s')v_{j}(s')\label{acoeff}
\end{eqnarray}

In evaluating the path integrals in Eq.~(\ref{Jpre}), we note that the function $\sigma(s)$ is linear in the effective action $S_{eff}[\sigma,\delta]$. Hence, the result of doing the path integral over $\sigma(s)$ gives a functional delta-function which basically enforces $\delta(s)$ to obey the equation of motion in Eq.~(\ref{cleq2}). However, the boundary conditions for $\delta(s)$ are $\delta(\tau_{i})=\delta(\tau_{f})=0$, the only allowed solution would be $\delta(s)=0$. As a result, the path integrals over $\sigma(s)$ and $\delta(s)$ give a function of $\tau_{i}$ and $\tau_{f}$, ${\cal Z}[\tau_{t},\tau_{i}]$, which is independent of $\Delta_{cl}$. 

One can obtain ${\cal Z}$ by imposing the normalization of the reduced density matrix, ${\rm Tr}(\rho_{R})=1$. From Eq.~(\ref{rhoR}), we can see that this normalization condition requires the evolution operator to satisfy
\begin{eqnarray}
    \int_{-\infty}^{\infty}d\Sigma_{f}\,{\cal J}_{R}(\Sigma_{f},0,t_{f}|\Sigma_{i},\Delta_{i},t_{i})=\delta(\Delta_{i})
    &\Rightarrow&{\cal Z}\int_{\infty}^{\infty}d\Sigma_{f}\ e^{-ib_{2}\Sigma_f\Delta_{i}-ib_{4}\Sigma_{i}\Delta_{i}-a_{11}\Delta_{i}^2}=\delta(\Delta_{i})\nonumber\\
    &\Rightarrow&{\cal Z}=\frac{b_{2}}{2\pi}
\end{eqnarray}
We therefore arrive at the final result in this derivation for the evolution operator
\begin{eqnarray}\label{evoope}
     &&{\cal J}_{R}(Q_{+f},Q_{-f},\tau_{f}|Q_{+ i},Q_{-i},\tau_{i})\nonumber\\
     &=&\left(\frac{b_{2}}{2\pi}\right)e^{ib_{1}\Sigma_f\Delta_f -ib_{2}\Sigma_f\Delta_{i}+ib_{3}\Sigma_{i}\Delta_f-ib_{4}\Sigma_{i}\Delta_{i}}\,e^{-a_{11}\Delta_{i}^2-a_{12}\Delta_{i}\Delta_f-a_{22}\Delta_f^{2}}
\end{eqnarray}

Now we are in a position to evaluate the reduced density matrix at any time $\tau$. Suppose that initially $Q$ is in the ground state. Then its density matrix is given by
\begin{eqnarray}\label{inigro}
    \bar{\rho}_{R}^{(0)}(\Sigma_i,\Delta_i)=\sqrt{\frac{m\omega}{\pi}}\,e^{-\frac{m\omega}{4}\left(4\Sigma_{i}^{2}+\Delta_{i}^{2}\right)}
\end{eqnarray}
where we have used an overbar to indicate the time-independent density matrix of an harmonic oscillator.
The density matrix at time $\tau_{f}$ can be obtained from the evolution equation in Eq.~(\ref{rhoR}) with the form of the evolution operator in Eq.~(\ref{evoope}) by integrating over $\Sigma_{i}$ and $\Delta_{i}$. Since it is at most quadratic in these variables up in the exponential, the integrals can be done exactly. Hence, the final density matrix can be expressed as
\begin{eqnarray}\label{rhoRt}
    \rho_{R}^{(0)}(\Sigma_f,\Delta_f,\tau_{f})={\cal N}\,e^{-{\cal A}\,\Delta_{f}^{2}-i\,2\,{\cal B}\,\Delta_{f}\Sigma_{f}-{\cal C}\,\Sigma_{f}^{2}}
\end{eqnarray}
where
\begin{eqnarray}
    {\cal N}&=&\frac{b_{2}}{2\sqrt{\pi}}\left(\frac{b_{4}^{2}}{4m\omega}+\frac{m\omega}{4}+a_{11}\right)^{-1/2}\nonumber\\
    {\cal A}&=&-\frac{1}{4}\left(\frac{b_{4}^{2}}{4m\omega}+\frac{m\omega}{4}+a_{11}\right)^{-1}\left(\frac{b_{3}b_{4}}{2m\omega}-a_{12}\right)^{2}+\frac{b_{3}^{2}}{4m\omega}+a_{22}\nonumber\\
    {\cal B}&=&-\frac{b_{1}}{2}+\frac{b_{2}}{4}\left(\frac{b_{4}^{2}}{4m\omega}+\frac{m\omega}{4}+a_{11}\right)^{-1}\left(\frac{b_{3}b_{4}}{2m\omega}-a_{12}\right)\nonumber\\
    {\cal C}&=&\frac{b_{2}^{2}}{4}\left(\frac{b_{4}^{2}}{4m\omega}+\frac{m\omega}{4}+a_{11}\right)^{-1}
\end{eqnarray}
This density matrix will be crucial in our subsequent derivations of the expectation values of functions of $Q$ and its momentum $P$ with respect to time. Then, the corresponding transition probabilities from the ground state to the excited states will be expressed in terms of these expectation values.

\subsection{Expectation values and transition probabilities}
From the density matrix $\rho_{R}$ in Eq.~(\ref{rhoRt}), one can evaluate the expectation value of an operator ${\cal O}$ in the quantum state which is initially in the ground state (Eq.~(\ref{inigro})) by taking the trace ${\rm Tr}({\cal O}\rho_{R}^{(0)})$. Since the density matrix is gaussian, it is obvious that the expectation values $\langle Q\rangle$ and $\langle P\rangle$ vanish. On the other hand,
\begin{eqnarray}\label{q2exp}
    \langle Q^{2}\rangle_{f}^{(0)}&=&{\rm Tr}(Q^{2}\rho_{R}^{(0)}(\tau_{f}))\nonumber\\
    &=&\int_{-\infty}^{\infty}d\Sigma_{f}\,\Sigma_{f}^{2}\,\rho_{R}^{(0)}(\Sigma_{f},0,\tau_{f})\nonumber\\
    &=&\frac{1}{2{\cal C}}\nonumber\\
    &=&\frac{b_{4}^{2}}{2m\omega b_{2}^{2}}+\frac{m\omega}{2b_{2}^{2}}+\left(\frac{2}{b_{2}^{2}}\right)a_{11}
\end{eqnarray}
The subscript $f$ shows that the expectation value is evaluated at time $\tau_{f}$ and the superscript $(0)$ indicates that the quantum state at the initial time $\tau_{i}$ is in the ground state.

In a similar fashion the expectation values $\langle P^{2}\rangle_{f}^{(0)}$ and $\langle \{Q,P\}\rangle_{f}^{(0)}$ are given by
\begin{eqnarray}\label{p2exp}
    &&\langle P^{2}\rangle_{f}^{(0)}\nonumber\\
    &=&{\rm Tr}(P^{2}\rho_{R}^{(0)}(\tau_{f}))\nonumber\\
    &=&\int_{-\infty}^{\infty}d\Sigma_{f}\int_{-\infty}^{\infty}d\Delta_{f}\left(\frac{\partial^{2}}{\partial \Delta^{2}_{f}}\delta(\Delta_{f})\right)\rho_{R}^{(0)}(\Sigma_{f},\Delta_{f},\tau_{f})\nonumber\\
    &=&2\left({\cal A}+\frac{{\cal B}^{2}}{{\cal C}}\right)\nonumber\\
    &=&\left(\frac{1}{2m\omega b_{2}^{2}}\right)(b_{1}b_{4}-b_{2}b_{3})^{2}+\frac{m\omega b_{1}^{2}}{2b_{2}^{2}}
    +\left(\frac{2b_{1}^{2}}{b_{2}^{2}}\right)a_{11}+\left(\frac{2b_{1}}{b_{2}}\right)a_{12}+2a_{22}
\end{eqnarray}
and 
\begin{eqnarray}\label{qpexp}
    \langle \{Q,P\}\rangle_{f}^{(0)}&=&{\rm Tr}(\{Q,P\}\,\rho_{R}^{(0)}(\tau_{f}))\nonumber\\
    &=&(-i)\int_{-\infty}^{\infty}d\Sigma_{f}\int_{-\infty}^{\infty}d\Delta_{f}\left(-\frac{\partial}{\partial \Delta_{f}}\delta(\Delta_{f})\right)\left(\Sigma_{f}+\frac{\Delta_{f}}{2}\right)\rho_{R}^{(0)}(\Sigma_{f},\Delta_{f},\tau_{f})\nonumber\\
    &&\ \ +(-i)\int_{-\infty}^{\infty}d\Sigma_{f}\int_{-\infty}^{\infty}d\Delta_{f}\left(-\frac{\partial}{\partial \Delta_{f}}\delta(\Delta_{f})\right)\left(\Sigma_{f}-\frac{\Delta_{f}}{2}\right)\rho_{R}^{(0)}(\Sigma_{f},\Delta_{f},\tau_{f})\nonumber\\
    &=&-\frac{2{\cal B}}{{\cal C}}\nonumber\\
    &=&\left(\frac{b_{4}}{m\omega b_{2}^{2}}\right)(b_{1}b_{4}-b_{2}b_{3})+\frac{m\omega b_{1}}{b_{2}^{2}}
    +\left(\frac{4b_{1}}{b_{2}^{2}}\right)a_{11}+\left(\frac{2}{b_{2}}\right)a_{12}
\end{eqnarray}

The transition probabilities of the detector from one quantum state to the other can also be formulated using the density matrices. Suppose that the detector is in its ground state initially. It is described at the initial time $\tau_{i}$ by the density matrix $\bar{\rho}_{R}^{(0)}$ as given by Eq.~(\ref{inigro}). At time $\tau_{f}$, the quantum state of the detector evolves to $\rho_{R}^{(0)}(\tau_{f})$ in Eq.~(\ref{rhoRt}). The probability that the detector will be, for example, in the first excited state at time $\tau_{f}$ is given by ${\rm Tr}(\,\bar{\rho}_{R}^{(1)}\rho_{R}^{(0)}(\tau_{f}))$, where $\bar{\rho}_{R}^{(1)}$ is the density matrix of the first excited state of a harmonic oscillator as given by
\begin{eqnarray}
    \bar{\rho}_{R}^{(1)}(\Sigma_{f},\Delta_{f})=\sqrt{\frac{m^{3}\omega^{3}}{4\pi}}\,(4\Sigma_{f}^{2}-\Delta_{f}^{2})\,e^{-\frac{m\omega}{4}(4\Sigma_{f}^{2}+\Delta_{f}^{2})}
\end{eqnarray}
Therefore, this transition probability can be worked out exactly as
\begin{eqnarray}\label{trapro}
    P_{0\rightarrow 1}(\tau_{f})&=&\int_{-\infty}^{\infty}d\Sigma_{f}\int_{-\infty}^{\infty}d\Delta_{f}\ \bar{\rho}_{R}^{(1)}(\Sigma_{f},\Delta_{f})\ \rho_{R}^{(0)}(\Sigma_f,\Delta_f,\tau_{f})\nonumber\\
    &=&\frac{2\,\sqrt{m^3\omega^3\,{\cal C}}\ (4{\cal A}-{\cal C})}{[(4{\cal A}+m\omega)({\cal C}+m\omega)+4{\cal B}^2]^{3/2}}\nonumber\\
    &=&\left[\langle Q^{2}\rangle_{f}^{(0)} \langle P^2\rangle_{f}^{(0)}-\frac{1}{4}[\langle\{Q,P\}\rangle_{f}^{(0)}]^{2}-\frac{1}{4}\right]\times\nonumber\\
    &&\ \ \left\{\left[\langle P^2\rangle_{f}^{(0)}+\frac{m\omega}{2}\right]\left[\langle Q^{2}\rangle_{f}^{(0)}+\frac{1}{2m\omega}\right]-\frac{1}{4}\left[\langle\{Q,P\}\rangle_{f}^{(0)}\right]^{2}\right\}^{-3/2}
\end{eqnarray}
This is the expression we shall use in our subsequent section in the evaluation of the transition probabilities of the UD detector from the ground state to the first excited state under the influence of the impulsive plane wave.






\section{Application to impulsive plane wave spacetimes}

The impulsive plane wave space we are considering has a delta function profile. The corresponding metric is given by \cite{GV91}
\begin{eqnarray}
    ds^{2}=-2\,du\,dv+\sum_{a=1}^{2}\eta_{a}\,\delta(u)\,(x^{a})^{2}\,du^{2}+\sum_{a=1}^{2}(dx^{a})^2
\end{eqnarray}
where $\eta_{1}$ and $\eta_{2}$ are constants. Here $u$ and $v$ are the light-cone coordinates with $u=t-z$ and $v=(t+z)/2$, and the impulsive wave is located at $u=0$.  Away from this wave, that is, $u<0$ and $u>0$, the spacetime is Minkowski.

In this spacetime the only non-vanishing Ricci tensor component at the wavefront is $R_{uu}$ It is proportional to $\eta_{1}+\eta_{2}$. Hence, for $\eta_{1}+\eta_{2}=0$, we have a Ricci flat spacetime which can be regarded as a pure gravitational wave. We shall consider in some detail later for the case with $\eta_{1}=\eta=-\eta_{2}$. Due to the weak energy condition, we have $\eta\geq 0$. Another interesting case will be $\eta_{1}=\eta_{2}=-\eta$ in which the Weyl tensor vanishes and this corresponds to a pure electromagnetic wave.

For the detector, we assume that it is fixed at the origin with the trajectory $z^{\mu}=(u,u/2,\vec{0})$. As it is discussed in \cite{PM24}, the Wightman function remains invariant under the transformation of any inertial trajectory to $(u,u/2,\vec{0})$. Hence, our discussion here should also be valid for any other inertial trajectories.

\subsection{Wightman function, retarded Green function, and Hadamard function}
The quantum field theory in the impulsive plane wave spacetimes have been considered in \cite{Klimcik88,GV91}. For a massless scalar field $\Phi$, the corresponding Klein-Gordon equation in this spacetime can be written as
\begin{eqnarray}\label{KGeqn}
    \left[-2\,\frac{\partial}{\partial u}\frac{\partial}{\partial v}+\sum_{a=1,2}\frac{\partial}{\partial x^{a}}\frac{\partial}{\partial x^{a}}-\sum_{a=1,2}\eta_{a}\,\delta(u)(x^{a})^2\frac{\partial^{2}}{\partial v^{2}}\right]\Phi=0
\end{eqnarray}
The mode functions are defined as follows \cite{Klimcik88,GV91,Cho23}. For the so-called in-modes, the mode function is that of a Minkowski mode function
\begin{eqnarray}
    \Phi_{k_{-}\,\vec{k}}^{\rm in}(z)=N_{k_{-}}e^{-ik_{-}v}e^{-\frac{iu}{2k_{-}}}e^{i\vec{k}\cdot\vec{x}}
\end{eqnarray}
for $u<0$. Here the normalization constant $N_{k_{-}}=[(2\pi)^3(2k_{-})]^{-1/2}$ and $z^{\mu}=(u,v,\vec{x})$ is a general spacetime point. For $u>0$, due to the delta function term in Eq.~(\ref{KGeqn}), we have the extra factor
\begin{eqnarray}
    \Phi_{k_{-}\,\vec{k}}^{\rm in}(z)=N_{k_{-}}e^{-ik_{-}v}e^{i\vec{k}\cdot\vec{x}}\int\frac{d^{2}x'\,d^{2}k'}{(2\pi)^{2}}e^{-i(\vec{k}-\vec{k}')\cdot(\vec{x}-\vec{x}')}e^{-\frac{iu}{2k_{-}}}e^{\frac{i}{2}k_{-}\sum_{a=1,2}\eta_{a}(x^{a})^2}
\end{eqnarray}
Similarly, one can also define the out-mode $\Phi_{k_{-}\,\vec{k}}^{\rm out}(z)$ as having the Minkowski form for $u>0$ and for $u<0$
\begin{eqnarray}
    \Phi_{k_{-}\,\vec{k}}^{\rm out}(z)=N_{k_{-}}e^{-ik_{-}v}e^{i\vec{k}\cdot\vec{x}}\int\frac{d^{2}x'\,d^{2}k'}{(2\pi)^{2}}e^{-i(\vec{k}-\vec{k}')\cdot(\vec{x}-\vec{x}')}e^{-\frac{iu}{2k_{-}}}e^{-\frac{i}{2}k_{-}\sum_{a=1,2}\eta_{a}(x^{a})^2}
\end{eqnarray}
Since we assume that the detector interacts with the quantum modes in the Minkowski vacuum before encountering the impulsive wave, we shall use the in-modes to construct the various two-point functions.

First, we look at the Wightman function from the in-modes,
\begin{eqnarray}
    G_{+}(z,z')=\int_{0}^{\infty}dk_{-}\int d^{2}k\ \Phi_{k_{-}\,\vec{k}}^{\rm in}(z)\Phi_{k_{-}\,\vec{k}}^{\rm in\,*}(z')
\end{eqnarray}
One can find a detailed derivation of the Wightman function in \cite{Cho23}. We just quote the result here. For a massless scalar field,
\begin{eqnarray}\label{wightman}
    G_{+}(z,z')=\frac{\sqrt{\Delta(u,u')}}{8\pi^{2}}\left[\frac{1}{\sigma(z,z')+i(u-u')\epsilon}\right]
\end{eqnarray}
where we have used the $i\epsilon$ prescription. Here we have introduced two biscalars $\Delta(u,u')$ and $\sigma(z,z')$. The world function $\sigma(z,z')$ is related to the distance between two points $u$ and $u'$ along a geodesic, while the van Vleck determinant $\Delta(u,u')$ describes the focusing of geodesic flows. By analyzing these biscalars, one can have a better understanding of the properties of the spacetime. 

For both $0>u>u'$ and $u>u'>0$, the world function is in the Minkowski form
\begin{eqnarray}
    \sigma(z,z')=-(u-u')(v-v')+\frac{1}{2}\sum_{a=1,2}(x^{a}-x'^{a})^2
\end{eqnarray}
For $u>0$ and $u'<0$, we have \cite{Cho23}
\begin{eqnarray}
    &&\sigma(z,z')\nonumber\\
    &=&\frac{1}{2}(u-u')\Big\{-2(v-v')\nonumber\\
    &&\ \ +\sum_{a=1,2}(u-u'-uu'\eta_{a})^{-1}\left[(1-u'\eta_{a})(x^{a})^2+(1+u\eta_{a})(x'^{a})^{2}-2x^{a}x'^{a}\right]\Big\}
\end{eqnarray}
For the van Vleck determinant, we have for $0>u>u'$ and $u>u'>0$,
\begin{eqnarray}
    \Delta(u,u)=1
\end{eqnarray}
and for $u>0$ and $u'<0$,
\begin{eqnarray}\label{vanVle}
    \Delta(u,u')=(u-u')^2(u-u'-uu'\eta_{1})^{-1}(u-u'-uu'\eta_{2})^{-1}
\end{eqnarray}

For the trajectory of the detector specified by $z^{\mu}=(u,u/2,\vec{0})$, the Wightman function in Eq.~(\ref{wightman}) can be simplified to
\begin{eqnarray}
    G_{+}(u,u')&=&\frac{\sqrt{\Delta(u,u')}}{8\pi^{2}}\left[\frac{1}{-(u-u')^2/2+i(u-u')\epsilon}\right]\nonumber\\
    &=&\sqrt{\Delta(u,u')}\ G_{+}^{M}(u,u')
\end{eqnarray}
where $G_{+}^{M}(u,u')$ is the Wightman function for the same trajectory in the Minkowski spacetime. To proceed we can choose the most convenient representation of the Minkowski Wightman function for our purpose. As we shall see, the best choice is
\begin{eqnarray}
    G_{+\Lambda}^{M}(u,u')=\frac{1}{4\pi^{2}}\int_{0}^{\Lambda}\,dk\,k\,e^{-ik(u-u')}
\end{eqnarray}
so that the resulting integrals will be in much simpler forms. Here we have introduced a momentum cutoff $\Lambda$ to regularize the $k$ integral, where $\Lambda$ will be treated as the largest scale in the problem.

In the influence action in Eq.~(\ref{SIF}), there are two more two-point functions we need to consider. The first one is the dissipation kernel defined by Eq.~(\ref{disker}). This is necessary for the study of the classical equations of motion in Eqs.~(\ref{cleq1}) and (\ref{cleq2}). Using the representation of the Wightman function above, we can write
\begin{eqnarray}\label{disD}
    D_{\Lambda}(u,u')&=&i\,\theta(u-u')\sqrt{\Delta(u,u')}\left[G_{+\Lambda}^{M}(u,u')-G_{+\Lambda}^{M}(u',u)\right]\nonumber\\
    &=&\theta(u-u')\frac{\sqrt{\Delta(u,u')}}{2\pi^{2}}\int_{0}^{\Lambda}\,dk\,k\,\sin[k(u-u')]
\end{eqnarray}
The other two-point function is the noise kernel given by the Hadamard function
\begin{eqnarray}\label{noiseN}
    N_{\Lambda}(u,u')&=&\frac{1}{2}\left[G_{+\Lambda}^{M}(u,u')+G_{+\Lambda}^{M}(u',u)\right]\nonumber\\
    &=&\frac{\sqrt{\Delta(u,u')}}{4\pi^{2}}\int_{0}^{\Lambda}\,dk\,k\,\cos[k(u-u')]
\end{eqnarray}

\subsection{Equations of motion}
As we have seen in Sec.~II, the evaluation of the expectation values and transition probabilities relies crucially on the solutions $u_{1}$, $u_{2}$, $v_{1}$, and $v_{2}$ to the equations of motion in Eqs.~(\ref{cleq1}) and (\ref{cleq2}). Hence, we need to study these equations more closely. We treat these equations as a initial value problem. Since the impulsive plane wave spacetime possesses no Cauchy surface, one can regard the constant $u$ surfaces as a substitute and use $u$ as the affine parameter for evolution. 

Take $u_{i}$ as the initial parameter and $u$ as the final one. For $0>u>s>u_i$ and $u>s>u_i>0$, the van Vleck determinant $\Delta(s,s')=1$. Using the form of the dissipation kernel in Eq.~(\ref{disD}), the equation of motion in Eq.~(\ref{cleq1}) can be expressed as
\begin{eqnarray}\label{mineqn}
    &&\partial_{s}^{2}f(s)+\omega_{0}^{2}\,f(s)-\left(\frac{\lambda^{2}}{m}\right)\int_{u_{i}}^{u}ds'\,D_{\Lambda}(s,s')\,f(s')=0\nonumber\\
    &\Rightarrow&\partial_{s}^{2}f(s)+\omega_{0}^{2}\,f(s)-\left(\frac{\lambda^{2}}{2\pi^{2}m}\right)\int_{u_{i}}^{s}ds'\,f(s')\int_{0}^{\Lambda}dk\,k\,\sin[k(s-s')]=0
\end{eqnarray}
If we take the integration over $k$, the result will be
\begin{eqnarray}
    \int_{0}^{\Lambda}dk\,k\,\sin[k(s-s')]=(s-s')^{-2}\,\Big(\sin[\Lambda(s-s')]-\Lambda(s-s')\cos[\Lambda(s-s')]\Big)
\end{eqnarray}
which is a function that concentrates around $s'=s$ for large $\Lambda$. Therefore, to analyze Eq.~(\ref{mineqn}), we can Taylor-expand the function $f(s')$ around $s'=s$.
\begin{eqnarray}
    f(s')=f(s)-(s-s')\,f'(s)+\frac{1}{2}(s-s')^{2}f''(s)+\cdots
\end{eqnarray}
and the last term in Eq.~(\ref{mineqn}) becomes
\begin{eqnarray}
    &&-\left(\frac{\lambda^{2}}{2\pi^{2}m}\right)\int_{0}^{\Lambda}dk\,k\,\int_{u_{i}}^{s}ds'\,\sin[k(s-s')]\left[ f(s)-(s-s')\,f'(s)+\frac{1}{2}(s-s')^{2}f''(s)+\cdots\right]\nonumber\\
    &=&-\left(\frac{\lambda^{2}}{2\pi^{2}m}\right)\left[\Lambda\,f(s)-\frac{\pi}{2}f'(s)+\cdots\right]
\end{eqnarray}
where the ellipsis represent terms which are of the order of $1/\Lambda$ or higher. Putting this result into Eq.~(\ref{mineqn}), we have
\begin{eqnarray}
    &&\partial_{s}^{2}f(s)+\omega_{0}^{2}\,f(s)-\left(\frac{\lambda^{2}}{2\pi^{2}m}\right)\left[\Lambda\,f(s)+\frac{\pi}{2}f'(s)+\cdots\right]=0\nonumber\\
    &\Rightarrow&\partial_{s}^{2}f(s)+2\gamma\, f'(s)+\omega^{2} f(s)=0\label{cleqn}
\end{eqnarray}
where
\begin{eqnarray}
    \omega^{2}=\omega_{0}^{2}-\frac{\lambda^{2}\Lambda}{2\pi^{2}m}\hskip 20pt;\hskip 20pt \gamma=\frac{\lambda^{2}}{8\pi m}
\end{eqnarray}
Note that $\omega$ can be viewed as the renormalized frequency of the oscillator. This equation represents a damped harmonic oscillator with the damping constant $\gamma$.

Next, we examine the behavior of the equation across the wavefront, that is, $u_{i}<0$ and $u>0$. The equation becomes
\begin{eqnarray}\label{ppeqn}
    \partial_{s}^{2}f(s)+\omega_{0}^{2}\,f(s)-\left(\frac{\lambda^{2}}{2\pi^{2}m}\right)\int_{u_{i}}^{s}ds'\,f(s')\sqrt{\Delta(s,s')}\int_{0}^{\Lambda}dk\,k\,\sin[k(s-s')]=0
\end{eqnarray}
Since we are interested in understanding whether the presence of the front will affect the behaviors of the solutions to the equation, especially whether the solutions and their derivatives will be discontinuous across the front, we integrate the equation between $\epsilon\geq s\geq -\epsilon$ and take the limit $\epsilon\rightarrow 0$ afterwards. For the first two terms in Eq.~(\ref{ppeqn}), we have
\begin{eqnarray}
    \int_{-\epsilon}^{\epsilon}ds\left[\partial_{s}^{2}f(s)+\omega_{0}^{2}\,f(s)\right]=f'(0^{+})-f'(0^{-})
\end{eqnarray}
as $\epsilon\rightarrow 0$.

To integrate the last term in Eq.~(\ref{ppeqn}), we take without loss of generality that $u_{i}=-\epsilon$. Then, for small $s$ on both sides of the front, we assume that
\begin{eqnarray}
    f(s)&=&f(0^{+})+sf'(0^{+})+\frac{1}{2}s^{2}f''(0^{+})+\cdots\hskip 20pt {\rm for}\ s>0\nonumber\\
    f(s)&=&f(0^{-})+sf'(0^{-})+\frac{1}{2}s^{2}f''(0^{-})+\cdots\hskip 20pt {\rm for}\ s<0
\end{eqnarray}
Integrating between $\epsilon\geq s\geq -\epsilon$, we have the integrals
\begin{eqnarray}
    &&\int_{-\epsilon}^{\epsilon}ds\,\int_{-\epsilon}^{s}ds'\,f(s')\sqrt{\Delta(s,s')}\int_{0}^{\Lambda}dk\,k\,\sin[k(s-s')]\nonumber\\
    &=&\int_{0}^{\epsilon}ds\,\int_{0}^{s}ds'\,f(s')\int_{0}^{\Lambda}dk\,k\,\sin[k(s-s')]+\int_{-\epsilon}^{0}ds\,\int_{-\epsilon}^{s}ds'\,f(s')\int_{0}^{\Lambda}dk\,k\,\sin[k(s-s')]\nonumber\\
    &&\ \ +\int_{0}^{\epsilon}ds\,\int_{-\epsilon}^{0}ds'\,f(s')\sqrt{\Delta(s,s')}\int_{0}^{\Lambda}dk\,k\,\sin[k(s-s')]\nonumber\\
\end{eqnarray}
In the first two terms $\Delta(s,s')=1$ because both $s$ and $s'$ are on the same side of the front. For small $\epsilon$, these two terms give
\begin{eqnarray}
    \int_{0}^{\epsilon}ds\,\int_{0}^{s}ds'\,f(s')\int_{0}^{\Lambda}dk\,k\,\sin[k(s-s')]&=&\frac{\Lambda^{3}}{18}f(0^{+})\epsilon^{3}+\cdots\\
    \int_{-\epsilon}^{0}ds\,\int_{-\epsilon}^{s}ds'\,f(s')\int_{0}^{\Lambda}dk\,k\,\sin[k(s-s')]&=&\frac{\Lambda^{3}}{18}f(0^{-})\epsilon^{3}+\cdots
\end{eqnarray}
For the third term, the van Vleck determinant is given by Eq.~(\ref{vanVle}). Rescaling $s\rightarrow s\epsilon$ and $s'\rightarrow s'\epsilon$, we can then expand the integrand in power of $\epsilon$.
\begin{eqnarray}
    &&\int_{0}^{1}ds\,\int_{-1}^{0}ds'\,f(\epsilon s')\sqrt{\Delta(\epsilon s,\epsilon s')}\int_{0}^{\Lambda}dk\,k\,\sin[k\,\epsilon\,(s-s')]\nonumber\\
    &=&\int_{0}^{1}ds\,\int_{-1}^{0}ds'\int_{0}^{\Lambda}dk\left[\epsilon f(0^{-})k^{2}(s-s')+\frac{\epsilon^{2}}{2}\bigg(f(0^{-})(\eta_{1}+\eta_{2})s+2f'(0^{-})(s-s')\bigg)+\cdots\right]\nonumber\\
    &=&\frac{\Lambda^{3}}{3}f(0^{+})\,\epsilon+\cdots
\end{eqnarray}
As $\epsilon\rightarrow 0$, all these three terms go to zero. Consequently, the dissipation term in the equation will not introduce any discontinuities in the solutions and their derivatives as long as $\Lambda$ is large but finite. In this respect, the equation of motion in Eq.~(\ref{cleqn}) is therefore valid throughout the whole impulsive plane wave spacetime.

Since Eq.~(\ref{cleqn}) is just the equation of motion for a damped harmonic oscillator, the solutions are well known. For the boundary conditons in Eq.~(\ref{u12bc}), we have
\begin{eqnarray}\label{usol}
    u_{1}(s)=e^{-\gamma(s-u_{i})}\,\frac{\sin[\Omega(u-s)]}{\sin[\Omega(u-u_{i})]}\hskip 20pt;\hskip 20pt u_{2}(s)=e^{\gamma(u-s)}\,\frac{\sin[\Omega(s-u_{i})]}{\sin[\Omega(u-u_{i})]}
\end{eqnarray}
where $\Omega=\sqrt{\omega^2-\gamma^2}$. Here we consider only the underdamped case with $\omega>\gamma$ so $\Omega$ is real.

Similar analysis can be applied to the other equation of motion in Eq.~(\ref{cleq2}) which gives 
\begin{eqnarray}
    \partial_{s}^{2}f(s)-2\gamma\, f'(s)+\omega^{2} f(s)=0
\end{eqnarray}
The solutions with the boundary conditions in Eq.~(\ref{v12bc}) are
\begin{eqnarray}\label{vsol}
    v_{1}(s)=e^{\gamma(s-u_{i})}\,\frac{\sin[\Omega(u-s)]}{\sin[\Omega(u-u_{i})]}\hskip 20pt;\hskip 20pt v_{2}(s)=e^{-\gamma(u-s)}\,\frac{\sin[\Omega(s-u_{i})]}{\sin[\Omega(u-u_{i})]}
\end{eqnarray}

\subsection{Effects of the impulsive wave}
In this subsection we are interested in studying the effects of the impulsive plane wave on the quantum state of the detector. In particular, we shall investigate the difference in evolutions of the expectation values of operators as well as  transition probabilities between quantum states with and without the wave. 

First, we examine the expectation values of $Q^2$, $P^2$, and $\{Q,P\}$ as indicated in Eqs.~(\ref{q2exp}) to (\ref{qpexp}). We see that they are given in terms of the coefficients $b_{i}$ and $a_{ij}$ of the evolution operator in Eq.~(\ref{evoope}). The coefficients $b_{i}$ are related to the solutions $u_{1}$ and $u_{2}$ as indicated in Eq.~(\ref{bcoeff1}). In the impulsive plane wave spacetime here, these solutions have been constructed above in Eq.~(\ref{usol}). Hence, using these solutions, we have
\begin{eqnarray}\label{bcoeff2}
    &&b_{1}=m\Big(-\gamma+\Omega\,\cot[\Omega(u-u_{i})]\Big)\hskip 20pt;\hskip 20pt b_{2}=m\,e^{\gamma(u-u_{i})}\Omega\,\csc[\Omega(u-u_{i})]\nonumber\\
    &&b_{3}=-m\,e^{-\gamma(u-u_{i})}\Omega\,\csc[\Omega(u-u_{i})]\hskip 20pt;\hskip 20pt b_{4}=-m\Big(\gamma+\Omega\,\cot[\Omega(u-u_{i})]\Big)
\end{eqnarray}
They are basically oscillating functions with frequency $\Omega$. Due to the exponential factor $e^{\gamma(u-u_{i})}$ in $b_{2}$, this coefficient increases in amplitude with $u$. While the amplitude of $b_{3}$ decreases with $u$ due to the factor $e^{-\gamma(u-u_{i})}$. Note that the coefficients $b_{i}$ are independent of the presence of the wave. 

In the impulsive plane wave spacetime, the coefficients $a_{ij}$ as given in Eq.~(\ref{acoeff}) can be expressed as
\begin{eqnarray}
    a_{ij}&=&\frac{\lambda^{2}}{4\pi^{2}(1+\delta_{ij})}\Bigg\{\int_{0}^{u}ds\int_{0}^{u}ds'\int_{0}^{\Lambda}dk\,k\ v_{i}(s)\,v_{j}(s')\cos[k(s-s')]\nonumber\\
    &&\hskip 60pt +\int_{u_{i}}^{0}ds\int_{u_{i}}^{0}ds'\int_{0}^{\Lambda}dk\,k\ v_{i}(s)\,v_{j}(s')\cos[k(s-s')]\nonumber\\
    &&\hskip 60pt +\int_{0}^{u}ds\int_{u_{i}}^{0}ds'\int_{0}^{\Lambda}dk\,k\sqrt{\Delta(s,s')}\  v_{i}(s)\,v_{j}(s')\cos[k(s-s')]\Bigg\}
\end{eqnarray}
where we have used the noise kernel in Eq.~(\ref{noiseN}).
We can see that the effects of the wave are only present in the third term through the van Vleck determinant since $s$ and $s'$ are on opposite sides of the wave. If we take $\Delta(s,s')=1$ in the third term, we would have the Minkowski result. Therefore, to obtain the difference of the coefficients between with and without the wave, we can subtract out the Minkowski result.
\begin{eqnarray}\label{deltaa}
    \Delta a_{ij}&\equiv& a_{ij}-a_{ij}^{M}\nonumber\\
    &=&\frac{\lambda^{2}}{4\pi^{2}(1+\delta_{ij})}\int_{0}^{u}ds\int_{-1/\Lambda}^{0}ds'\int_{0}^{\Lambda}dk\,k\,\left(\sqrt{\Delta(s,s')}-1\right)\,  v_{i}(s)\,v_{j}(s')\cos[k(s-s')]\nonumber\\
\end{eqnarray}
with only the third term remaining.
Note that we have taken $u_i=-1/\Lambda$. The reason is the following. We assume that the detector is in the ground state initially when it encounters the wave. However, we cannot set $u_{i}$ to zero. In that case, we would have a vanishing result in Eq.~(\ref{deltaa}). To have the initial time close to zero, we use the smallest scale $1/\Lambda$ in the problem and set $u_{i}=-1/\Lambda$.

We can now express the expectation value differences with and without the wave in terms of these coefficient differences. From Eq.~(\ref{q2exp}), we have
\begin{eqnarray}
    \Delta\langle Q^{2}\rangle_{f}^{(0)}&=&\langle Q^{2}\rangle_{f}^{(0)}-\langle Q^{2}\rangle_{f}^{(0)M}\nonumber\\
    &=&\left(\frac{2}{b_{2}^{2}}\right)\Delta a_{11}
\end{eqnarray}
As shown in Eq.~(\ref{deltaa}), $\Delta a_{11}$ can be simplified using the solutions $v_{i}$ given in Eq.~(\ref{vsol}) and the van Vleck determinant in Eq.~(\ref{vanVle}). Furthermore, with the coefficient $b_{2}$ in Eq.~(\ref{bcoeff2}),
\begin{eqnarray}\label{DQ2}
    &&\Delta\langle Q^{2}\rangle_{f}^{(0)}\nonumber\\
    &=&\Delta\langle Q^{2}\rangle_{i}^{(0)}\left(\frac{8\omega\gamma}{\pi\Omega^{2}}\right)e^{-2\gamma u}\int_{0}^{u}ds\int_{-1/\Lambda}^{0}ds'\,e^{\gamma(s+s')}\sin[\Omega(u-s)]\sin[\Omega(u-s')]\nonumber\\
    &&\hskip 80pt (s-s')^{-2}\Big[(s-s')(s-s'-ss'\eta_{1})^{-1/2}(s-s'-ss'\eta_{2})^{-1/2}-1\Big]\nonumber\\
    &&\hskip 100pt\Big[-1+\cos[\Lambda(s-s')]+\Lambda(s-s')\sin[\Lambda(s-s')]\Big]
\end{eqnarray}
where $\langle Q^{2}\rangle_{i}^{(0)}=1/2m\omega$ is the initial expectation value of $Q^2$ in the ground state. Note that we have already evaluated the integral over $k$.


Similarly, for the expectation value $\langle P^{2}\rangle_{f}^{(0)}$ in Eq.~(\ref{p2exp}), we have
\begin{eqnarray}\label{DP2}
    &&\Delta\langle P^{2}\rangle_{f}^{(0)}\nonumber\\
    &=&\langle P^{2}\rangle_{f}^{(0)}-\langle P^{2}\rangle_{f}^{(0)M}\nonumber\\
    &=&\left(\frac{2b_{1}^{2}}{b_{2}^{2}}\right)\Delta a_{11}+\left(\frac{2b_{1}}{b_{2}}\right)\Delta a_{12}+2\Delta a_{22}\nonumber\\
    &=&\langle P^{2}\rangle_{i}^{(0)}\left(\frac{8\gamma}{\pi\Omega}\right)e^{-2\gamma u}\int_{0}^{u}ds\int_{-1/\Lambda}^{0}ds'\,e^{\gamma(s+s')}\left(\frac{\gamma}{\Omega}\sin[\Omega(u-s)]-\cos[\Omega(u-s)]\right)\nonumber\\
    && \left(\frac{\gamma}{\Omega}\sin[\Omega(u-s')]-\cos[\Omega(u-s')]\right)\Big[-1+\cos[\Lambda(s-s')]+\Lambda(s-s')\sin[\Lambda(s-s')]\Big]\nonumber\\
    &&(s-s')^{-2}\Big[(s-s')(s-s'-ss'\eta_{1})^{-1/2}(s-s'-ss'\eta_{2})^{-1/2}-1\Big]
\end{eqnarray}
where $\langle P^{2}\rangle_{i}^{(0)}=m\omega/2$ is the initial expectation value of $P^2$ in the ground state. For the expectation value $\langle \{Q,P\}\rangle_{f}^{(0)}$  in Eq.~(\ref{qpexp}),
\begin{eqnarray}\label{DQP}
    &&\Delta\langle \{Q,P\}\rangle_{f}^{(0)}\nonumber\\
    &=&\left(\frac{4b_{1}}{b_{2}^{2}}\right)\Delta a_{11}+\left(\frac{2}{b_{2}}\right)\Delta a_{12}\nonumber\\
    &=&\left(\frac{4\gamma}{\pi\Omega}\right)e^{-2\gamma u}\int_{0}^{u}ds\int_{-1/\Lambda}^{0}ds'\,e^{\gamma(s+s')}
    \Bigg\{\sin[\Omega(u-s)]\cos[\Omega(u-s')]\nonumber\\
    &&\ \ \ \ +\cos[\Omega(u-s)]\sin[\Omega(u-s')]
    -\left(\frac{2\gamma}{\Omega}\right)\sin[\Omega(u-s)]\sin[\Omega(u-s')]\Bigg\}\nonumber\\
    &&\ \  (s-s')^{-2}\Big[(s-s')(s-s'-ss'\eta_{1})^{-1/2}(s-s'-ss'\eta_{2})^{-1/2}-1\Big]\nonumber\\
    &&\ \ \ \ \Big[-1+\cos[\Lambda(s-s')]+\Lambda(s-s')\sin[\Lambda(s-s')]\Big]
\end{eqnarray}


Finally, we come to the transition probability of the detector from the ground state to the first excited state due to the effect of the impulsive wave. In Eq.~(\ref{trapro}), it can be evaluated from the expectation values $\langle Q^{2}\rangle_{f}^{(0)}$, $\langle P^{2}\rangle_{f}^{(0)}$, and $\langle \{Q,P\}\rangle_{f}^{(0)}$. Again, the effect of the wave can be obtained by subtracting out the contribution from the Minkowski value. Hence, for the transition probability
\begin{eqnarray}
    \Delta P_{0\rightarrow 1}=P_{0\rightarrow 1}-P_{0\rightarrow 1}^{M}
\end{eqnarray}
In the case that the expectation values in Minkowski spacetime are much larger than the effects due to the wave, that is, $\langle {\cal O}\rangle^{M}\gg \Delta\langle{\cal O}\rangle$, one can express $\Delta P_{0\rightarrow 1}$ to first power of the expectation value differences,
\begin{eqnarray}\label{DP01}
    \Delta P_{0\rightarrow 1}&=&\frac{1}{2}\left\{\left[\langle P^2\rangle_{f}^{(0)M}+\frac{m\omega}{2}\right]\left[\langle Q^{2}\rangle_{f}^{(0)M}+\frac{1}{2m\omega}\right]-\frac{1}{4}\left[\langle\{Q,P\}\rangle_{f}^{(0)M}\right]^{2}\right\}^{-5/2}\nonumber\\
    &&\Bigg\{2\left[\left(\langle P^2\rangle_{f}^{(0)M}+\frac{m\omega}{2}\right)\left(\langle Q^{2}\rangle_{f}^{(0)M}+\frac{1}{2m\omega}\right)-\frac{1}{4}\left(\langle\{Q,P\}\rangle_{f}^{(0)M}\right)^{2}\right]\nonumber\\
&&\ \ \left[\langle Q^{2}\rangle_{f}^{(0)M}\Delta\langle P^{2}\rangle_{f}^{(0)}+\langle P^{2}\rangle_{f}^{(0)M}\Delta \langle Q^{2}\rangle_{f}^{(0)}-\frac{1}{2}\langle\{Q,P\}_{f}^{(0)M}\Delta\langle\{Q,P\}\rangle_{f}^{(0)}\right]\nonumber\\
    &&\ -3\left[\langle Q^{2}\rangle_{f}^{(0)M} \langle P^2\rangle_{f}^{(0)M}-\frac{1}{4}[\langle\{Q,P\}\rangle_{f}^{(0)M}]^{2}-\frac{1}{4}\right]\nonumber\\
    &&\ \ \Bigg[\left(\langle P^2\rangle_{f}^{(0)M}+\frac{m\omega}{2}\right)\Delta\langle Q^{2}\rangle_{f}^{(0)}+\left(\langle Q^{2}\rangle_{f}^{(0)M}+\frac{1}{2m\omega}\right)\Delta\langle P^2\rangle_{f}^{(0)}\nonumber\\
    &&\ \ \ \ -\frac{1}{2}\langle\{Q,P\}\rangle_{f}^{(0)M}\Delta\langle\{Q,P\}\rangle_{f}^{(0)}\Bigg]\Bigg\}+\cdots
\end{eqnarray}
In the next section, we shall use the formulas in Eqs.~(\ref{DQ2}) to (\ref{DQP}) and also Eq.~(\ref{DP01}) to calculate the expectation values and the transition probabilities with and without the impulsive wave in some explicit cases.

\section{Specific examples: Degenerate and non-degenerate cases}
In this section we present results on the response of the UD detector in two different impulsive plane wave spacetimes. The first one is the nondegenerate case with $\eta_{1}\neq\eta_{2}$. More specifically, we take $\eta_{1}=-\eta_{2}=\eta$. Since $\eta_{1}+\eta_{2}=0$, the spacetime is Ricci-flat and represents a pure gravitational wave. The second one is the degenerate case with $\eta_{1}=\eta_{2}=-\eta$ in which the Weyl tensor vanishes, corresponding to a null electromagnetic wave. We set $\omega=1$ as the reference scale and the cutoff scale to $\Lambda=100$.

\subsection{The gravitational wave}
\begin{figure}
\centering
	\begin{minipage}[t]{0.3\linewidth}
		\centering
        \includegraphics[width=4.5cm]{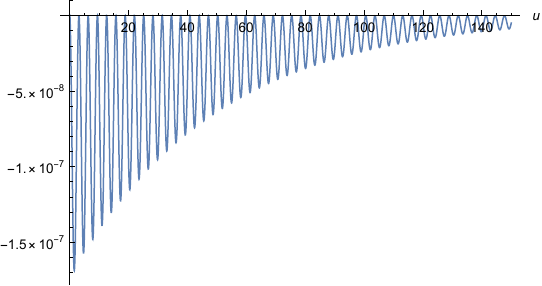}
	    \includegraphics[width=4.5cm]{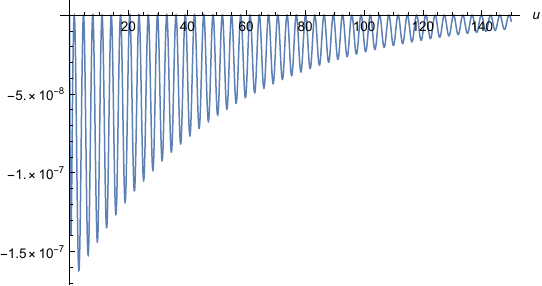}
        \includegraphics[width=4.5cm]{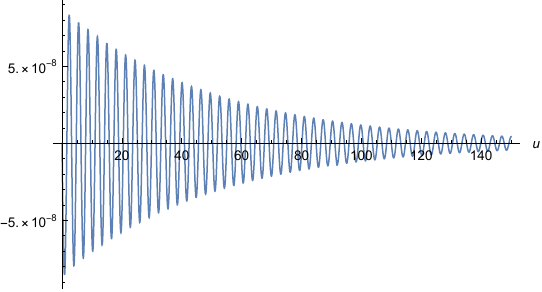}
        \subcaption{$\gamma=0.01,\,\eta_{1}=-\eta_{2}=1$}
	\end{minipage}
	\begin{minipage}[t]{0.3\linewidth}
		\centering
		\includegraphics[width=4.5cm]{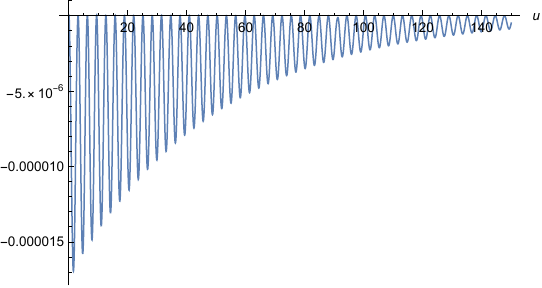}
        \includegraphics[width=4.5cm]{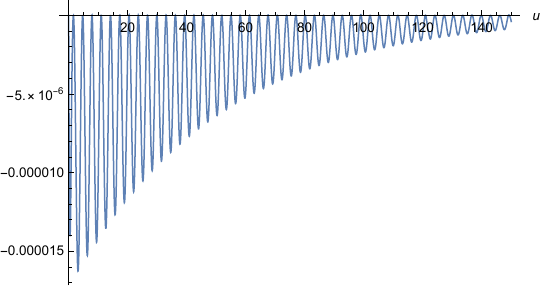}
        \includegraphics[width=4.5cm]{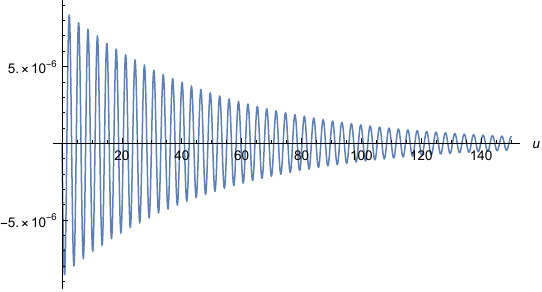}
        \subcaption{$\gamma=0.01,\,\eta_{1}=-\eta_{2}=10$}
	\end{minipage}
	\begin{minipage}[t]{0.3\linewidth}
		\centering
        \includegraphics[width=4.5cm]{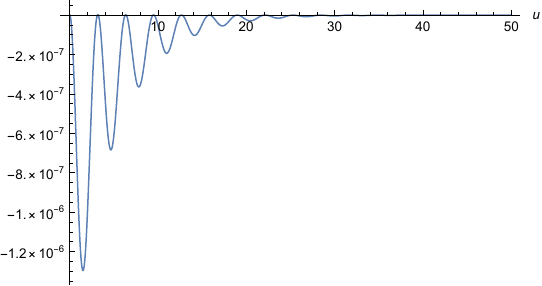}
        \includegraphics[width=4.5cm]{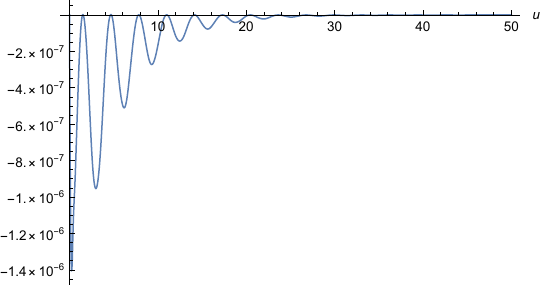}
        \includegraphics[width=4.5cm]{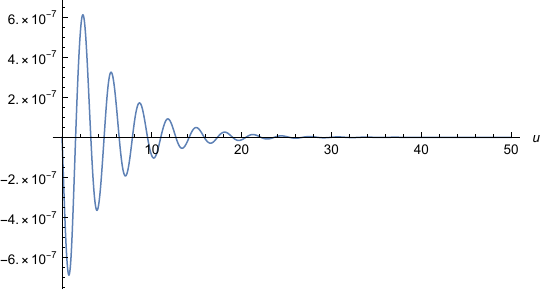}
        \subcaption{$\gamma=0.1,\,\eta_{1}=-\eta_{2}=1$}
	\end{minipage}
\caption{Effects of the gravitational wave on the expectation values $\Delta\langle Q^{2}\rangle_{f}^{(0)}/\langle Q^{2}\rangle_{i}^{(0)}$ (top panel), $\Delta\langle P^2\rangle_{f}^{(0)}/\langle P^2\rangle_{i}^{(0)}$ (middle panel) and $\Delta\langle \{Q,P\}\rangle_{f}^{(0)}$ (bottom panel).}\label{figexpln}	
\end{figure}

In this gravitational wave or non-degenerate case, we first examine the expectation values of $\Delta \langle Q^2\rangle$, $\Delta \langle P^2\rangle$, and $\Delta\langle\{Q,P\}\rangle$ and their evolutions. We perform the integrations numerically over $s$ and $s'$ as in Eqs.~(\ref{DQ2}) to (\ref{DQP}). The results are shown in Fig.~\ref{figexpln}. 
In the first column we have the evolutions of $\Delta\langle Q^{2}\rangle_{f}^{(0)}/\langle Q^{2}\rangle_{i}^{(0)}$, $\Delta\langle P^2\rangle_{f}^{(0)}/\langle P^2\rangle_{i}^{(0)}$, and $\Delta\langle \{Q,P\}\rangle_{f}^{(0)}$, respectively for the dissipation coefficient $\gamma=0.01$ and the strength of the wave $\eta_{1}=-\eta_{2}=1$. To see the trend of the evolution we consider first the top graph. Initially at $u_{i}=-1/\Lambda=-0.01$, the detector is in the ground state and the expectation value of $Q^{2}$ is $\langle Q^{2}\rangle_{i}^{(0)}=1/2m\omega$. Interaction with the quantum fields excites the detector and the expectation value will increase. However, encountering the impulsive plane wave at $u=0$ must have a suppressing effect on the excitation in such a way that the difference $\Delta\langle Q^{2}\rangle_{f}^{(0)}$ becomes negative. It oscillates basically with the frequency $2\Omega=2\sqrt{\omega^2-\gamma^2}\sim 2$, and the amplitude decreases with a decay constant related to $\gamma=0.01$. $\Delta\langle P^2\rangle_{f}^{(0)}/\langle P^2\rangle_{i}^{(0)}$ has an evolution almost identical to that of $\Delta\langle Q^{2}\rangle_{f}^{(0)}/\langle Q^{2}\rangle_{i}^{(0)}$. In contrast, $\Delta\langle \{Q,P\}\rangle_{f}^{(0)}$ oscillates about the $x$-axis as in the Minkowski case.

In the second column of Fig.~\ref{figexpln}, we show the expectation values again for $\gamma=0.01$, but with $\eta$ increased to 10. We can see that the overall trends with $u$ are similar to case (a). It is apparent that as the strength of the gravitational plane wave increases compared to case (a), the expectation values increase accordingly. The oscillation amplitude of the expectation values in case (a) is on the order of $10^{-7}$, whereas in case (b) it is on the order of $10^{-5}$. On the other hand, in the third column, the dissipation coefficient $\gamma$ is increased to 0.1, ten times larger than in case (a). Since $\gamma=\lambda^{2}/8\pi m$, a value of $\gamma$ comparable to $\omega$ corresponds to a strong coupling regime. Consequently, the detector response to the plane wave is heightened: the oscillation amplitudes are on the order of $10^{-6}$ (ten times larger than in case (a)), even though the frequency $2\Omega\sim 1.99$ remains nearly unchanged. Furthermore, because the decay rate is proportional to $\gamma$, the amplitudes decay much faster than in cases (a) and (b).

\begin{figure}
\centering
	\begin{minipage}[t]{0.3\linewidth}
		\centering
        \includegraphics[width=4.5cm]{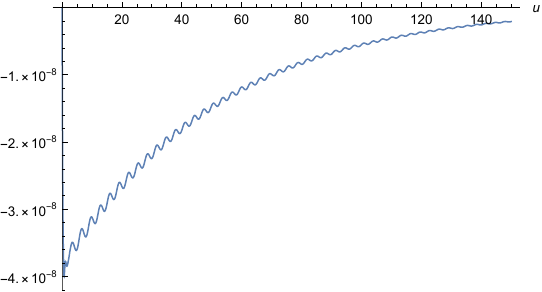}
        \subcaption{$\gamma=0.01,\,\eta_{1}=-\eta_{2}=1$}
	\end{minipage}
    \begin{minipage}[t]{0.3\linewidth}
		\centering
		\includegraphics[width=4.5cm]{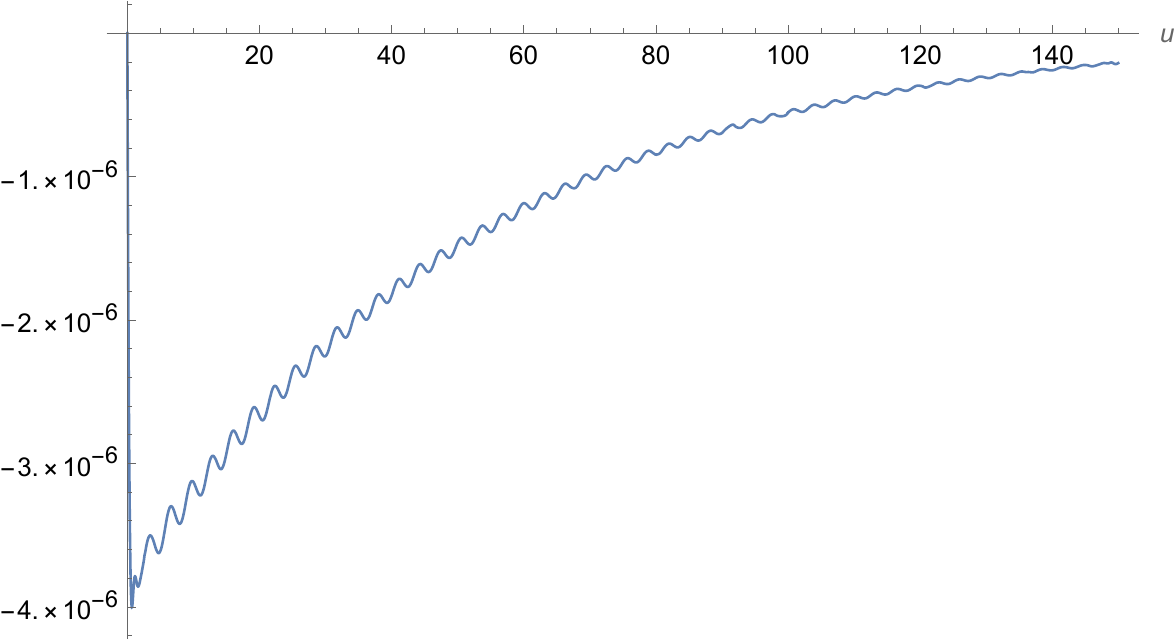}
        \subcaption{$\gamma=0.01,\,\eta_{1}=-\eta_{2}=10$}
	\end{minipage}
	\begin{minipage}[t]{0.3\linewidth}
		\centering
        \includegraphics[width=4.5cm]{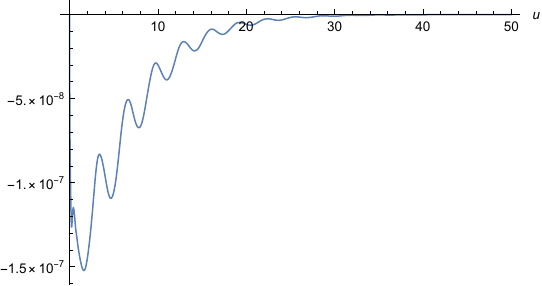}
        \subcaption{$\gamma=0.1,\,\eta_{1}=-\eta_{2}=1$}
	\end{minipage}
\caption{Effects of the gravitational wave on the transition probabilities $\Delta P_{0\rightarrow 1}$ from the ground state to the first excited state of the detector.}\label{figDPln}	
\end{figure}

The evolutions of the transition probabilities $\Delta P_{0\rightarrow 1}$ are shown in Fig.~\ref{figDPln}. From the results in Fig.~\ref{figexpln}, the magnitude of $\Delta\langle{\cal O}\rangle_{f}^{(0)}$ is at most on the order of $10^{-5}$, which is much smaller than $\langle{\cal O}\rangle_{f}^{(0)}$ (expected to be on the order of unity \cite{HH19a}). Therefore, we can use Eq.~(\ref{DP01}) to calculate these transition probabilities. For $\gamma=0.01$ and $\eta=1$, the result is shown in Fig.~\ref{figDPln}(a), where the suppressing effect reaches its maximum after encountering the wave and subsequently decays at a rate governed by $\gamma$. In Fig.~\ref{figDPln}(b), where the wave strength is increased to $\eta=10$, the trend remains identical to case (a), but with the peak magnitude increasing to the order of $10^{-6}$ compared to $10^{-8}$ in (a). When $\gamma$ is increased to $0.1$ in case (c), the amplitude jumps to a maximum on the order of $10^{-7}$, but it decays at a much faster rate proportional to $\gamma$. Consequently, there is a crossover in the value of $\Delta P_{0\rightarrow 1}$ between case (c) and case (a) around $u=10$.

\subsection{The electromagnetic wave}

\begin{figure}
\centering
	\begin{minipage}[t]{0.3\linewidth}
		\centering
        \includegraphics[width=4.5cm]{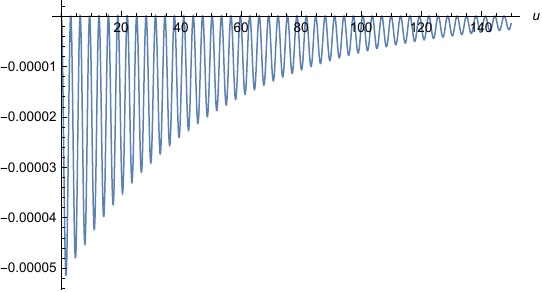}
	    \includegraphics[width=4.5cm]{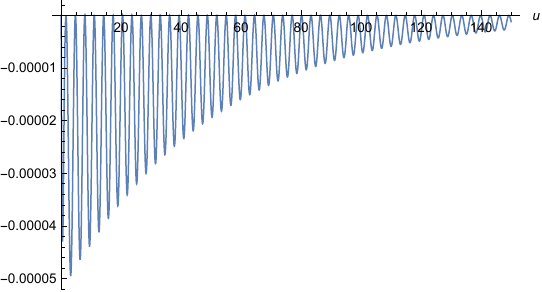}
        \includegraphics[width=4.5cm]{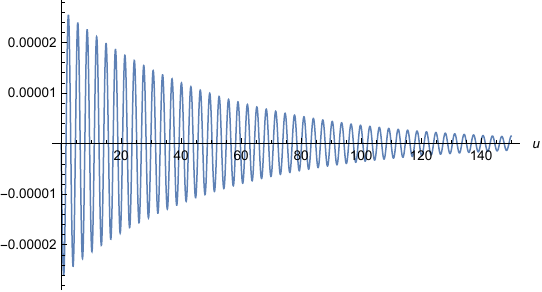}
        \subcaption{$\gamma=0.01,\,\eta_{1}=\eta_{2}=1$}
	\end{minipage}
	\begin{minipage}[t]{0.3\linewidth}
		\centering
		\includegraphics[width=4.5cm]{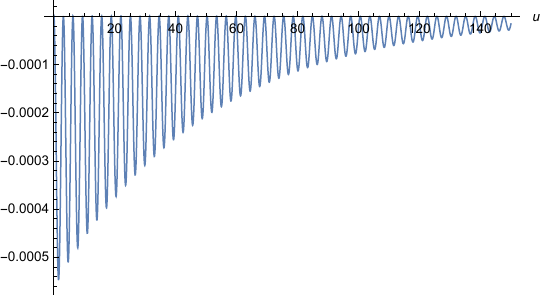}
        \includegraphics[width=4.5cm]{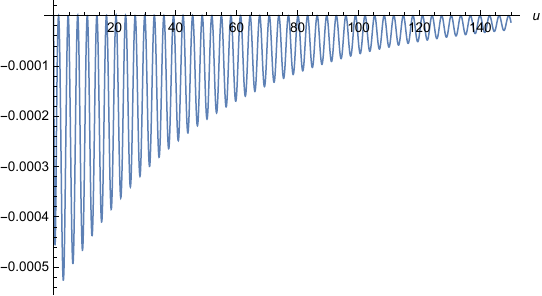}
        \includegraphics[width=4.5cm]{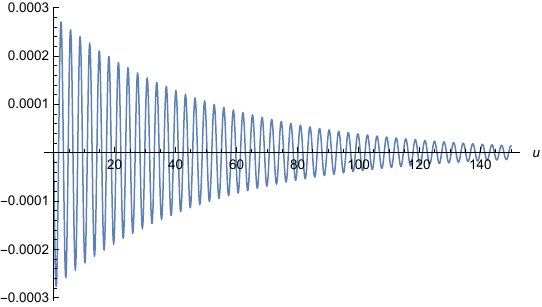}
        \subcaption{$\gamma=0.01,\,\eta_{1}=\eta_{2}=10$}
	\end{minipage}
	\begin{minipage}[t]{0.3\linewidth}
		\centering
        \includegraphics[width=4.5cm]{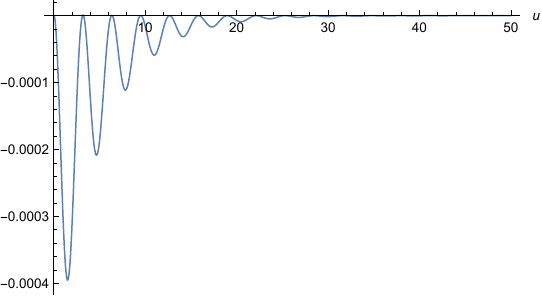}
        \includegraphics[width=4.5cm]{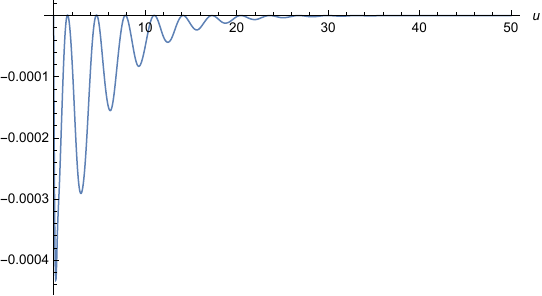}
        \includegraphics[width=4.5cm]{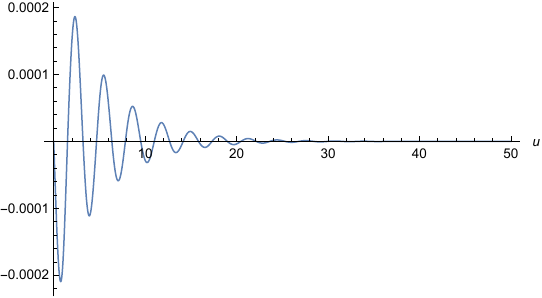}
        \subcaption{$\gamma=0.1,\,\eta_{1}=\eta_{2}=1$}
	\end{minipage}
\caption{Effects of the electromagnetic wave on the expectation values $\Delta\langle Q^{2}\rangle_{f}^{(0)}/\langle Q^{2}\rangle_{i}^{(0)}$ (top panel), $\Delta\langle P^2\rangle_{f}^{(0)}/\langle P^2\rangle_{i}^{(0)}$ (middle panel) and $\Delta\langle \{Q,P\}\rangle_{f}^{(0)}$ (bottom panel).}\label{figexplln}
\end{figure}
Here, we consider the degenerate case with $\eta_{1}=\eta_{2}=-\eta$, which represents a null electromagnetic plane wave. The resulting effects on the evolution of the expectation values after the detector encounters the wave are shown in Fig.~\ref{figexplln}. As in Fig.~\ref{figexpln}, the presence of the impulsive wave suppresses these expectation values. Furthermore, the functional dependencies on $u$ mirror those in the gravitational case, as both the oscillation frequencies and decay constants are identical. The only difference compared to the gravitational wave case lies in the magnitude of the oscillations. For example, in Fig.~\ref{figexpln}(a), the first peak of $\Delta\langle\{Q,P\}\rangle_{f}^{(0)}$ is on the order of $10^{-7}$, whereas the corresponding expectation value in Fig.~\ref{figexplln}(a) is on the order of $10^{-5}$. This indicates that an electromagnetic wave exerts a stronger effect on the detector quantum state than a gravitational wave with the same parameter $\eta$.

\begin{figure}
\centering
	\begin{minipage}[t]{0.3\linewidth}
		\centering
        \includegraphics[width=4.5cm]{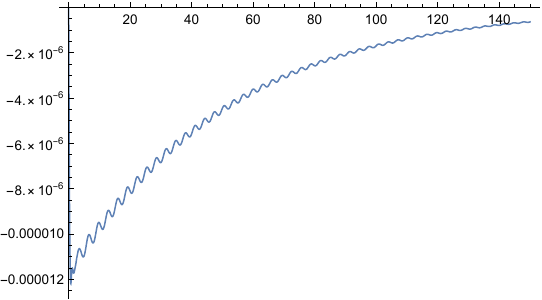}
        \subcaption{$\gamma=0.01,\,\eta_{1}=\eta_{2}=1$}
	\end{minipage}
    \begin{minipage}[t]{0.3\linewidth}
		\centering
		\includegraphics[width=4.5cm]{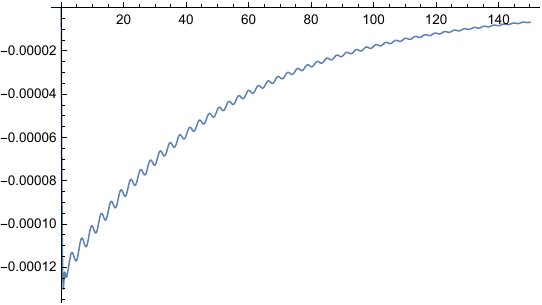}
        \subcaption{$\gamma=0.01,\,\eta_{1}=\eta_{2}=10$}
	\end{minipage}
	\begin{minipage}[t]{0.3\linewidth}
		\centering
        \includegraphics[width=4.5cm]{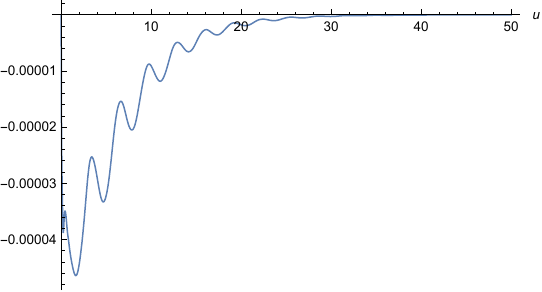}
        \subcaption{$\gamma=0.1,\,\eta_{1}=\eta_{2}=1$}
	\end{minipage}
\caption{Effects of the electromagnetic wave on the transition probabilities $\Delta P_{0\rightarrow 1}$ from the ground state to the first excited state of the detector.}\label{figDPlln}	
\end{figure}

In Fig.~\ref{figDPlln}, a similar conclusion can be drawn. Compared to Fig.~\ref{figDPln}, we see that the evolution of the transition probabilities $\Delta P_{0\rightarrow 1}$ during the interaction with the impulsive electromagnetic wave follows the same trend as in the gravitational case. However, the electromagnetic wave also exerts a stronger suppressive effect on the transition probabilities than a gravitational wave with the same $\eta$. In Fig.~\ref{figDPlln}(a), the minimum value is on the order of $10^{-5}$, whereas for the corresponding graph in Fig.~\ref{figDPln}(a), the value is only on the order of $10^{-8}$.

\section{Conclusions and discussions}

Following the open quantum system approach, in particular the influence functional formalism, we have considered the response of a UD detector interacting with a massless scalar field in impulsive plane wave spacetimes. The internal structure of the detector is modeled as a harmonic oscillator. By subtracting the Minkowski effect, we obtain the expectation values of $Q^2$, $P^2$, and $\{Q,P\}$, as well as the transition probabilities from the ground state to the first excited state due to the influence of the impulsive plane wave. We have explicitly calculated two cases: one with a vanishing Ricci tensor ($\eta_1 = -\eta_2$), corresponding to a pure gravitational wave, and the other with a vanishing Weyl tensor ($\eta_1 = \eta_2$), corresponding to a null electromagnetic wave. In both cases, we see that the presence of the wave has a suppressing effect on the transition of the UD detector to the first excited state. Indeed, one possible extension of our work is to investigate whether this is a general rule or if it holds only for particular values of $\eta$.

Since our approach is non-perturbative, our results are valid across all values of the coupling strength. Therefore, we have considered cases with $\gamma=0.01$ as well as $\gamma=0.1$. Note that $\gamma=\lambda^2/8\pi m$, so different values of $\gamma$ correspond to different values of the coupling constant $\lambda$. This contrasts with the results in \cite{GKMTT21} and \cite{PM24}, where the response of the UD detector was also studied. However, those studies utilize a perturbative approach, which is predominant in the literature on UD detector theory. As shown in \cite{LH07,HH19a}, this perturbative approach is insufficient for investigating detector responses in the strong coupling regime or for exploring late-time behaviors. With the framework developed in this paper, it is possible to study UD detector responses non-perturbatively in either impulsive plane wave spacetimes or sandwich waves with finite extent.

An intriguing feature of the impulsive plane wave is the presence of conjugate planes, near which quantities such as the Wightman function diverge. For our choice of initial point $u_{i}=-1/\Lambda$ close to the wave location at $u=0$, no conjugate planes are present. However, if we choose an earlier initial $u_i$ for state preparation, conjugate planes may appear in the region $u>0$. Therefore, in future work, we can consider a more general $u_{i}$ to investigate how conjugate planes affect the quantum state of the detector, particularly its transition probabilities to excited states.

Another possible extension of our work is to consider decoherence dynamics. Suppose the initial quantum state of the detector is a pure state with non-zero off-diagonal elements in its density matrix. Due to interaction with the environmental quantum field, decoherence occurs, leading to the decay of these off-diagonal terms over time. We can then investigate whether the presence of an impulsive or sandwich plane wave enhances or suppresses this decay. Furthermore, our framework can be extended to multi-detector systems to explore detector-detector entanglement. By evaluating the time evolution of entanglement measures such as negativity or concurrence for the joint reduced density matrix, we can quantify the impact of the passing wave on quantum entanglement.


\begin{acknowledgments}
The author would like to thank Jen-Tsung Hsiang and Bei-Lok Hu for helpful discussions especially on the open quantum system approach to UD detectors. This work was supported in part by the National Science and
Technology Council (NSTC) of Taiwan, Republic of
China, under Grant Nos. MOST 114-2112-M-032-007 and MOST 115-2112-M-032-005.
\end{acknowledgments}



\end{document}